\documentclass[letterpaper,useAMS,usenatbib]{mn2e}

\usepackage{times}
\usepackage[dvips]{graphicx}
\usepackage[totalwidth=480pt, totalheight=680pt]{geometry}

\newcommand{\etal}{{\rm et al.}~}

\newcommand{\be}{\begin{equation}}
\newcommand{\ee}{\end{equation}}

\newcommand{\te}{\theta}

\newcommand{\bfl}{{\mathbf l}}

\newcommand{\spose}[1]{\hbox to 0pt{#1\hss}}
\newcommand{\lta}{\mathrel{\spose{\lower 3pt\hbox{$\mathchar"218$}}
 \raise 2.0pt\hbox{$\mathchar"13C$}}}
\newcommand{\gta}{\mathrel{\spose{\lower 3pt\hbox{$\mathchar"218$}}
 \raise 2.0pt\hbox{$\mathchar"13E$}}}

\begin{document}

\author{Micha{\l} J.~Chodorowski}

\title{A direct consequence of the expansion of space?}
\date{\small \it Copernicus Astronomical Center, Bartycka 18, 00--716 
Warsaw, Poland }
\maketitle

\begin{abstract}
\normalsize 
Consider radar ranging of a distant galaxy in a Friedman-Lema{\^\i}tre
cosmological model. In this model the comoving coordinate of the
galaxy is constant, hence the equations of null geodesics for photons
travelling to the distant galaxy and back imply
\begin{displaymath}
\int_{\tau_e}^{\tau_r} \frac{{\rm d}\tau}{a(\tau)} =
\int_{\tau_r}^{\tau_o} \frac{{\rm d}\tau}{a(\tau)} \,.
\end{displaymath}
Here, $\tau_e$, $\tau_r$ and $\tau_o$ are respectively the times of
emission, reflection and observation of the reflected photons, and
$a(\tau)$ is the scale factor. Since the universe is expanding,
$a(\tau)$ is a monotonically increasing function, so the return travel
time, $\tau_o - \tau_r$, must be greater than the forward travel time,
$\tau_r - \tau_e$. Clearly, space expands, and on their way back, the
photons must travel a longer distance! The present paper explains why
this argument for the expansion of space is wrong. We argue that,
unlike the expansion of the cosmic substratum, the expansion of space
is unobservable. We therefore propose to apply to it -- just like to
the ether -- Ockham's razor.
\end{abstract}

\section{Introduction}
\label{sec:intro}
Modern cosmology is based upon general relativity (GR), and many
results of GR defy our expectations based upon special relativity
(SR). For example, in Friedman-Lema{\^\i}tre (FL) cosmological models
distant galaxies recede with superluminal recession velocities (e.g.,
Davis \& Lineweaver 2004), and the distance to the particle horizon is
greater than $c\, \tau_0$, where $\tau_0$ is the age of the
Universe. To `explain' these and other GR effects in cosmology, the
idea of the Expansion of Space (EoS) is evoked. Namely, in this {\em
interpretation\/} of the FL models, the motion of galaxies is not a
normal motion through space, but instead space itself is expanding and
carrying the galaxies and other matter along. This description seems
to be supported by the existence of comoving coordinates, which remain
fixed for all particles of the FL cosmic substratum. The notion of the
EoS is intended to help understanding that, in cosmology, one should
{\em not\/} expect SR to hold. With respect to this particular
purpose, this notion certainly does its job. In our opinion, however, it
constitutes simultaneously a serious conceptual pitfall, on several
levels:

\begin{itemize}
\item On a philosophical level, it suggests that the expansion of the
 universe can be detached from the matter that is participating in the
 expansion. However, we know that, as he was constructing GR, Einstein
 was greatly influenced by the thoughts of German physicist and
 philosopher Ernst Mach. In the words of Rindler (1977), for Mach
 ``space is not a `thing' in its own right; it is merely an abstraction
 from the totality of distance-relations between matter". Therefore,
 the idea of expanding space `in its own right' is very much contrary
 to the spirit of GR.

\item On a physical level, it suggests that the EoS is a geometric
  effect, so space itself is absolute. Then, though abolished in SR,
  in cosmology absolute space reenters triumphally the cosmic arena,
  endowed with an additional attribute: expansion.

\item Again on a physical level, it suggests the existence of a new
mysterious force. If so, one can expect non-standard effects also on
small scales. For example, one might expect particles to be dragged
along by the EoS. Davis, Lineweaver and Webb (2003), Whiting (2004),
Barnes \etal (2006) and Peacock (2007) show that this is not the
case.\footnote{In the words of Davis \etal (2003), "the example of an
expanding universe in which an untethered galaxy approaches us exposes
the common fallacy that 'expanding space' is in some sense trying to
drag all points apart (\ldots) It does (\ldots) highlight the common
false assumption of a force or drag associated with the EoS."} Also,
wavelengths of laboratory photons should change roughly by the factor
$1 + H_0 \tau_{\rm exp}$, where $\tau_{\rm exp}$ is the duration of a
given experiment and $H_0$ is the Hubble constant (Lieu \& Gregory
2006a). This is also wrong (Lieu \& Gregory 2006b).

\item On a psychological level, it suggests that we cannot use our
  classical intuition. In other words, we cannot think of the physical
  problem as real, relativistic motions of matter in presence of
  gravity, but as an abstract, geometric effect, where no visual
  `models' are possible. However, according to Longair (2003), model
  building in physics is very important and useful: ``(\ldots) when I
  think about some topic in physics or astrophysics, I generally have
  some picture, or model, in my mind rather than an abstract or
  mathematical idea.''  Classical intuition certainly has its
  limitations -- in particular it fails for the Planck era of the
  cosmological expansion -- but here we are talking about {\em
  classical\/} GR and its effects.
\end{itemize}

One may argue that the concept of expanding space does have an
appealing visualization: the surface of an inflated balloon, with dots
on it representing galaxies. However, when interpreting this picture
as an illustration of the EoS, there is a problem. Really moving
galaxies have kinetic energy; do so those entirely driven by the
expansion of massless space? The answer is not clear, the more that
the latter are often claimed to be `effectively' at rest, i.e.,
relative to the cosmic microwave background. Therefore, while the
interpretation of the cosmological expansion as real relative motions
of the cosmic substratum leads naturally -- for nonrelativistic
velocities -- to the Newtonian interpretation of the FL equations (in
particular, energy conservation; Milne 1934; McCrea \& Milne 1934),
that based on the idea of expanding space does not.

Still, isn't space expanding from a global point of view? Spatial
sections of a closed FL model are three-spheres, whose radius of
curvature increases as $a(\tau)$. Here, $a(\tau)$ is the so-called
scale-factor, a universal function of cosmic time which describes how
the distances between all elements of the cosmic substratum (or,
fluid) grow with time. Therefore, the proper volume of a closed FL
universe increases as $[a(\tau)]^3$; more and more space thus
appears. The same is true for open and critical FL models, though in
these models the universe has infinite extent. Peacock (2007) calls
this effect ``global Expansion of Space''. He makes a clear
distinction between the concepts of global and local EoS, and attacks
only the latter: ``the very idea that the motion of distant galaxies
could affect {\em local\/} dynamics is profoundly anti-relativistic:
the equivalence principle says that we can always find a tangent frame
in which physics is locally special relativity''. Isn't the {\em
global\/} EoS a fact?

A consistent Machian (like the author of the present paper) would
reply that since the cosmic substratum is expanding, it is not
surprising that the volume occupied by it increases. However, it is
the cosmic substratum that is really expanding, Indeed, through
observations we can only detect the motions of the particles of the
cosmic substratum. For example, if we lived long enough, we could
measure the radar distance to a distant galaxy and discover that it
increases with time.\footnote{By sending photons to the galaxy at
several instants of time and noticing that the differences between the
arrival times of the reflected photons are greater than the
differences between the emission times.} Changing distance implies
motion, and increasing distance implies recession. But how can one
measure distance to something that is not material? We believe, after
Mach and Einstein, that `unobservables' -- like ether or its
cosmological disguise, expanding space -- have no place in physics. On
the other hand, it is true that the Big Bang was not an explosion in a
preexisting void. Galaxies do not move {\em through\/} space or {\em
in\/} space. In a Machian view, they move instead {\em with\/} space:
they simply enable space {\em to exist}.

Superluminal recession velocities of distant galaxies are used as an
argument for the EoS (Lineweaver \& Davis 2005). Specifically, it is
argued that the motions cannot be `normal', otherwise they would
violate SR. If we define `normal' motion as the one taking place in
Minkowskian spacetime, the last inference is correct. As mentioned
above, superluminality of distant galaxies and `acausality' of the
particle horizon are GR effects and they don't comply with
SR. However, by no means this implies that the only possible
interpretation of the GR cosmological equations is the
EoS!\footnote{Contrary to what is widely believed.} An alternative
interpretation, advocated here, is that these equations describe
nothing more than real relative motions of the particles of the cosmic
substratum. In this interpretation, superluminality of distant
galaxies and `acausality' of the particle horizon can be well
understood (though understanding always demands more effort than
deriving). In particular, in at least one FL model (namely, the empty
model), superluminality of distant galaxies is merely a coordinate
effect: the {\em inertial\/} recession velocities are subluminal
(Davis 2004, Chodorowski 2007, Gr{\o}n \& Elgar{\o}y 2007; see also
Chodorowski 2005).

The misunderstanding of identifying EoS with GR and real motions with
SR dates back to Milne (1934). For he wrote: ``The phenomenon of the
expansion of the universe has usually been discussed by students of
relativity by means of the concept of `expanding space'. This concept,
though mathematically significant, has by itself no physical content;
it is merely the choice of a particular mathematical apparatus for
describing and analysing phenomena. An alternative procedure is to
choose a static space, as in ordinary physics, and analyse the
expansion-phenomenon as actual motions in this space. (\ldots)
Einstein's general relativity adopts the first procedure; in my recent
treatment of the cosmological problem I adopted the second
procedure. (\ldots) the second procedure has the advantage that it
employs the space commonly used in physics.'' In his original work on
kinematic cosmology, Milne (1933) specified what he meant as `the
space commonly used in physics': ``flat, infinite, static Euclidean
space''. He also wrote: ``Moving particles in a static space will give
the same observable phenomena as stationary particles in `expanding'
space'' (Milne 1934). These statements are wrong in general. In his
desire to abolish `expanding space', Milne went too far: he attempted
to abolish GR, or to prove that it is {\em not\/} indispensable in
cosmology. This erroneous, dichotomic way of thinking about motions in
cosmology: either EoS and GR, or real motions and SR, has been
inherited and is shared by many contemporary authors. For example,
Abramowicz \etal (2006) show that FL cosmological models are not
(except for the Milne model) compatible with SR, and use this fact as
an argument for the EoS (see also Gr{\o}n \& Elgar{\o}y 2007). 

In the present paper, we argue against the EoS, but {\em not\/}
against GR. Specifically, in the framework of GR we analyze critically
another argument for the EoS, which is fairly often heard: the travel
time of photons. In all expanding FL universes, the forward travel
time of photons travelling to a distant galaxy and back is claimed to
be smaller than the return travel time. The difference in these two
travel times is commonly attributed to the phenomenon of the EoS. In
Section~\ref{sec:travel}, we relate the two travel times in the
conventional Robertson--Walker (RW) coordinates. We study in
Section~\ref{sec:empty} the special case of an empty universe and
compare the two times in the Minkowskian coordinates. In
Section~\ref{sec:non-empty}, we introduce {\em conformally\/}
Minkowskian coordinates for general FL models and obtain the resulting
relation between the two times. We discuss the travel-time effect and
some other effects commonly attributed to the EoS in
Section~\ref{sec:disc}. A summary is given in Section~\ref{sec:summ}.

\section{Robertson-Walker coordinates}
\label{sec:travel}
The metric of a homogeneous and isotropic universe is given by the 
RW line element:
\begin{equation}
ds^2 = c^2 d\tau^2 - a^2(\tau)[dx^2 + R_0^2 S^2(x/R_0) d\psi^2] .
\label{eq:RW}
\end{equation}
Here, 
\be
d\psi^2 = d\theta^2 + \sin^2\theta d\phi^2 ,
\label{eq:dOmega}
\end{equation}
and $R_0^{-2}$ is the present curvature of the universe.  The function
$S(x)$ equals $\sin(x)$, $x$, and $\sinh(x)$ for a closed, flat, and
open universe, respectively. The scale factor $a(\tau)$ relates the
physical, or proper, coordinates of a galaxy, $\bfl$, to its fixed or
comoving coordinates, $\mathbf{x}$: $\bfl = a \mathbf{x}$. This
function accounts for the expansion of the universe; its detailed time
dependence is determined by the FL equations.

Photons propagate along null geodesics, $ds = 0$. Let us place an
observer at the origin of the coordinate system; then the geodesic of
the photons emitted by the observer towards any distant galaxy will be
radial. We denote the comoving radial coordinate of a distant galaxy
by $x_g$. From the metric~(\ref{eq:RW}) we have
\begin{equation}
\int_{\tau_e}^{\tau_r} \frac{c\, {\rm d}\tau}{a(\tau)} = \!\int_0^{x_g}
{\rm d} x = x_g ,
\label{eq:forward} 
\end{equation}
where $\tau_e$ is the emission time of the photons and $\tau_r$ is the time
they reach the distant galaxy. Let's assume that at the distant galaxy
the photons are instantaneously reflected towards the observer. Since
the comoving coordinate of the distant galaxy is constant, we can
write an analogous equation for the returning photons. As a result,
\be 
\int_{\tau_e}^{\tau_r} \frac{{\rm d}\tau}{a(\tau)} =
\int_{\tau_r}^{\tau_o} \frac{{\rm d}\tau}{a(\tau)},
\label{eq:both}
\ee where $\tau_o$ is the observation time of the photons by the
observer at the origin. Now, $a(\tau)$ is a monotonically increasing
function, so for the two above integrals to be equal, $\tau_r$ must be
smaller than $(\tau_e + \tau_o)/2$. In other words, the return travel
time, $\tau_o - \tau_r$, must be greater than the forward travel time,
$\tau_r - \tau_e$. Eureka! Space expands, so on their way back, the
photons must cover a longer distance!
This is in a marked contrast with SR, where the two travel times are
always equal.

Perhaps surprisingly, there is a loophole in the above line of
reasoning. Namely, the instants of time $\tau_e$, $\tau_r$ and
$\tau_o$ are not measured in the same rest frame. While $\tau_e$ and
$\tau_o$ are measured in the observer's rest-frame, $\tau_r$ is
measured in the rest frame of the distant galaxy.\footnote{For in the
RW coordinates, time $\tau$ is always the time of a {\em local\/} rest
frame.} From now on, we will denote this latter time $\tau_r'$ rather
than $\tau_r$. These two frames -- the observer and the galaxy -- are
in relative motion. Therefore, due to relativity of simultaneity, in
the observer's frame the time attributed to the event of reflection of
the photons will be different ($\tau_r$ instead of $\tau_r'$). If both
rest-frames were globally inertial then the Lorentz transform would
relate them and one could predict the relation between $\tau_r$ and
$\tau_r'$ using solely SR. However, in the presence of the
gravitational field all inertial frames are only local. Therefore, the
only cosmological model in which we can relate the two instants of
time using SR is the empty model. In an empty universe there is no
gravity, so inertial frames {\em are\/} global. In the following
section we will calculate $\tau_r$ for this model.

\section{Empty universe}
\label{sec:empty}

Expansion of an empty universe is kinematic, $a(\tau) \propto
\tau$. Applied to Equation~(\ref{eq:both}), this yields
\be
\tau_r' = \sqrt{\tau_e \, \tau_o} \,. 
\label{eq:tau_r'}
\ee 
In the empty model the reflection time $\tau_r'$ (measured in the rest
frame of the distant galaxy) is thus the geometric mean of the
instants $\tau_e$ and $\tau_o$; it is indeed smaller than the
arithmetic mean. The dynamics of an empty universe can be described
entirely by means of the Milne kinematic model. In this model, the
arena of all cosmic events is pre-existing Minkowski spacetime. In
the origin of the coordinate system, $O$, at Minkowskian time $t = 0$
an `explosion' takes place, sending radially so-called Fundamental
Observers (FOs) with constant velocities in the range of speeds
$(0,c)$. The FO, which remains at the origin (the `central observer')
measures time $\tau = t$. (The adopted global Minkowskian time $t$ is
thus the proper time of the central observer). At time $\tau_e$ this
observer emits photons, which at time $\tau_r$ reach a FO riding on a
galaxy receding with velocity $v$, such that $v \tau_r = c (\tau_r -
\tau_e)$. Hence, \be v = (1 - \tau_e/\tau_r) c .
\label{eq:v}
\ee 
Relative to the set of synchronized clocks of the inertial frame of
the central observer, the clock carried out by the FO riding on the
distant galaxy, $\tau'$, delays: $\tau = \gamma(v) \tau'$. Here,
$\gamma(v) = (1-v^2/c^2)^{-1/2}$. In particular,
\be 
\tau_r = \gamma(v) \tau_r'.
\label{eq:tau_r}
\ee 
Using Equation~(\ref{eq:v}) in Equation~(\ref{eq:tau_r}) yields
\be
\tau_r = \frac{1}{2} \left(\frac{\tau_r'^{\,2}}{\tau_e} + \tau_e\right) .
\label{eq:tau_r2}
\ee
In turn, using Equation~(\ref{eq:tau_r'}) in
Equation~(\ref{eq:tau_r2}) yields finally
\be
\tau_r = \frac{\tau_e + \tau_o}{2} .
\label{eq:tau_r3}
\ee 
According to the central observer, the travel times $\tau_o - \tau_r$
and $\tau_r - \tau_e$ are thus equal. We see that in the empty model,
the effect of non-equal travel times is explicable {\em entirely\/} by
the special-relativistic phenomenon of time dilation. 

Reasoning that the effect can be attributed to the EoS was in fact
non-relativistic and as such had to be wrong, for it had to fail for
light propagation. This reasoning used the analogy of a swimmer
swimming in a river against the stream, so it implicitly assumed
non-relativistic, Galilean law of addition of velocities. Just after
photons' reflection, their local velocity (relative to the distant
galaxy) is $v' = c$, and the galaxy recedes from the observer with
velocity $V$. Using the Galilean law, the velocity of the reflected
photons relative to the observer is $v = v' - V = c - V < c$. Then, it
indeed takes longer time for the photons to return to the
observer.\footnote{More specifically, $\tau_o - \tau_r = (1 -
V/c)^{-1} (\tau_r - \tau_e)$. This Galilean relation can be obtained
alternatively by noticing that relative to the distant galaxy, the
reflected photons travel longer distance than $r_r = c(\tau_r -
\tau_e)$: this distance is $r_o = r_r + V (\tau_o - \tau_r)$. Then the
relation follows from the equality $r_o = c (\tau_o - \tau_r)$.}
However, one of the main postulates of theory of relativity is that if
$v' = c$, then also $v = c$. This postulate is a direct consequence of
the null results of all the ether-drift experiments
(Michelson--Morley, Kennedy--Thorndike, etc.). These experiments
attempted to indirectly detect the ether (identified with Absolute
Space, i.e., an absolute standard of rest) by detecting the motion of
the laboratory frame relative to it, assuming the Galilean law of
addition of velocities. Their null results imply that for light, the
analogy of a swimmer in a river does not work: its $v$ is $c$ in every
inertial frame. One cannot detect the flow of ether; similarly one
cannot detect the Hubble flow of expanding space. Einstein used
Ockham's razor to the ether -- as unobservable -- and declared it
non-existent. Are the concepts of ether and expanding space much
different?

\section{General Friedman-Lema{\^I}tre models}
\label{sec:non-empty}
In Milne's terminology, RW coordinates constitute `public space',
while inertial (Minkowskian) coordinates define `private space' of an
observer. In a non-empty universe there are no global inertial
frames. From this, Gr{\o}n \& Elgar{\o}y (2007) deduce that in the
real universe it is impossible to define `private space' of an
observer: ``In curved spacetime (\ldots) the only real space is the
public space.'' But this is not so.

All FLRW models are conformally flat. In other words, their line
elements can be expressed as a product of the Minkowski metric,
$ds^2_M = c^2 dt^2 - dr^2 - r^2 d\psi^2$, and a function of time and
distance:
\be
ds^2 = f^2(t,r) ds^2_M \,. 
\label{eq:conform}
\ee 
Relativists are aware of this fact: it is easy to show that the Weyl
tensor of the RW metric (Eq.~\ref{eq:RW}) identically vanishes
(Krasi{\'n}ski, private communication). Some modern textbooks on
cosmology do mention this (e.g.\ Peacock 1999). To our knowledge,
however, none of them presents an explicit form of the RW metric in
conformally Minkowskian coordinates. We think that this form is worth
reminding; it was derived and thoroughly discussed in a superb paper of
Infeld \& Schild (1945). The paper of Infeld \& Schild is highly
mathematical, so in our derivation below we will follow a more
intuitive approach outlined in Landau \& Lifshitz (1979).

If a universe is spatially flat, finding conformally Minkowskian
coordinates is trivial. We define the conformal time, $\tilde\eta$, by the
equation $d\tau = a d\tilde\eta$. Then the RW metric (Eq.~\ref{eq:RW})
becomes 
\be
ds^2 = a^2(\tilde\eta) (c^2 d\tilde\eta^2 - dx^2 - x^2 d\psi^2) \,,
\label{eq:RW_flat}
\ee
so it is conformally flat. However, here we are interested in finding
conformally Minkowskian coordinates for {\em all\/} FLRW models. Let
us concentrate on the case $\Omega_0 < 1$, where $\Omega_0$ is the
present value of the mean total energy density in the universe in
units of the critical density. (The case $\Omega_0 > 1$ can be treated
similarly). The radius of the curvature is
\be
R_0 = \frac{c H_0^{-1}}{\sqrt{1 - \Omega_0}} \,.
\label{eq:curv}
\ee
We rescale the comoving coordinate, $\chi \equiv x/R_0$, and define
the time-like coordinate $\eta$ by the equation $c d\tau = R_0 a
d\eta$. The RW metric then becomes
\be
ds^2 = R_0^2 a^2(\eta)[d\eta^2 - d\chi^2 - \sinh^2(\chi) d\psi^2] \,,
\label{eq:RW_eta}
\ee
so $\eta$ is {\em not\/} the conformal time. Let us now introduce new
variables 
\be
r = A{\rm e}^\eta \sinh\chi \,,
\label{eq:r}
\ee
and 
\be
c t = A{\rm e}^\eta \cosh\chi \,,
\label{eq:t}
\ee
where $A$ is a constant. The old variables expressed in terms of
the new ones are:
\be
A {\rm e}^\eta = \sqrt{c^2 t^2 - r^2} \,,
\label{eq:eta}
\ee
and
\be
\tanh\chi = \frac{r}{ct} \,.
\label{eq:chi}
\ee
From Equations~(\ref{eq:eta})--(\ref{eq:chi}) we obtain
\be
d\eta = \frac{c^2 t\, dt - r\,dr}{c^2 t^2 - r^2} \,,
\label{eq:d_eta}
\ee
and
\be
d\chi = \frac{c t\, dr - r c\,dt}{c^2 t^2 - r^2} \,.
\label{eq:d_chi}
\ee
This yields
\be
d\eta^2 - d\chi^2 = \frac{c^2 dt^2 - dr^2}{c^2 t^2 - r^2} \,.
\label{eq:diff}
\ee
From Equations~(\ref{eq:r}) and~(\ref{eq:eta}), $\sinh^2(\chi) =
r^2/(c^2 t^2 - r^2)$. Hence, 
\be 
d\eta^2 - d\chi^2 - \sinh^2(\chi) d\psi^2 = \frac{c^2 dt^2 - dr^2
- r^2 d\psi^2}{c^2 t^2 - r^2} \,,
\label{eq:interv}
\ee
or, using Equation~(\ref{eq:RW_eta}), 
\be 
ds^2 = \frac{R_0^2 a^2(\eta)}{c^2 t^2 - r^2} ds^2_M \,.
\label{eq:confor}
\ee
In the above equation, $\eta$ is a function of $t$ and $r$ given by
Equation~(\ref{eq:eta}).

Equation~(\ref{eq:confor}) is valid for {\em any\/} open FLRW
model. Here, for illustrative purposes we will restrict ourselves to a
matter-dominated open universe without the cosmological constant:
$\Omega_\Lambda = 0$, $\Omega_0 = \Omega_m < 1$. Then the scale factor
is
\be
a(\eta) = a_\star(\cosh\eta - 1) \,,
\label{eq:scale}
\ee
where 
\be
a_\star = \frac{\Omega_0}{2(1 - \Omega_0)} \,.
\label{eq:norm}
\ee 
The scale factor is normalized so that its present value $a_0 =
a(\eta_0) = 1$. We have $\cosh\eta - 1 = (1 - {\rm e}^\eta)^2/(2 {\rm
e}^\eta)$, and after some algebra, the square root of the conformal 
factor is

\be
\frac{R_0 a(\eta)}{\sqrt{c^2 t^2 - r^2}} = \frac{R_0 a_\star}{2 A}
\left(1 - \frac{A}{\sqrt{c^2 t^2 - r^2}} \right)^2 .
\label{eq:conf_fact}
\ee 
Adopting $A = a_\star R_0/2$, and using Equations~(\ref{eq:curv})
and~(\ref{eq:norm}), we obtain finally the RW metric in the form given
by Equation~(\ref{eq:conform}), with 
\be
f(t,r) = \left[1 - \frac{c H_0^{-1} \Omega_0}{4 (1 - \Omega_0)^{3/2} 
    \sqrt{c^2 t^2 - r^2}} \right]^2 .
\label{eq:conf_fact2}
\ee 
Note that in the limit $\Omega_0 \to 0$ the conformal factor $f(t,r)
\to 1$ and the metric~(\ref{eq:conform}) tends to the Minkowski
metric, as expected.\footnote{In the limit $\Omega_0 \to 1^{-}$ the
conformal factor (Eq.~\ref{eq:conf_fact2}) blows up, but then $R_0 \to
\infty$ and the metric tends to the form given by
Equation~(\ref{eq:RW_flat}).} The coordinates $t$ and $r$ are thus a
generalization of Minkowskian coordinates for non-empty
universes. Therefore, they naturally define `private space' of {\em
the\/} Fundamental Observer, on whom the metric is centered. More
specifically, we define his `private space' as space-like sections of
the metric~(\ref{eq:conform}), hypersurfaces $t = \rm const$. This
`private space' has many interesting properties, but their full
discussion is outside of the scope of the present paper and will be
given elsewhere. Let us only note in passing that from
Equation~(\ref{eq:chi}) FOs, or galaxies on which they ride, have
world-lines given by the equation
\be
r = \tanh(x/R_0) ct \,.
\label{eq:world_line}
\ee
Since the comoving coordinate $x$ of any galaxy is constant, all
galaxies recede from the central one with constant coordinate
velocities, with speed $v = \tanh(x/R_0) c$, or
\be 
\beta \equiv
\frac{v}{c} = \tanh(x/R_0) \,.
\label{eq:speed}
\ee 
From metric~(\ref{eq:conform}) we immediately see that in the
conformally Minkowskian coordinates, light-cones ($ds = 0$) are the
same as in SR. As a result, in these coordinates the speed of light is
exactly $c$. From Equation~(\ref{eq:speed}) it follows that the
recession velocities of all galaxies, even arbitrarily distant (in the
sense $x \to \infty$) are subluminal. This is in contrast to the
recession velocities in the RW coordinates, which become superluminal
for sufficiently distant galaxies. 

Let us now return to the main topic of the present paper, that is the
travel time of photons. Since in the conformally Minkowskian
coordinates light-cones are the same as in SR, in these coordinates
the travel times of photons travelling to a distant galaxy and back
are always equal. We thus see that both the superluminality of distant
galaxies and the light travel-time effect are merely coordinate
effects: they can be eliminated by the choice of a suitable coordinate
system. 

\section{Discussion}
\label{sec:disc}

An effect which is coordinate-independent is a real phenomenon, which
different observers will agree on. Such a phenomenon is the expansion
of the universe: one can describe it in many coordinate systems, but
the expansion scalar of the cosmic substratum, $\te \equiv
v^\mu_{;\mu}$ (here, the semicolon denotes a covariant derivative), is
an invariant and its measured value is positive. Conversely, an effect
which is entirely coordinate-dependent is not a real physical
phenomenon. It is solely an artifact of the used coordinate
system. The superluminality of distant galaxies and the light
travel-time effect belong to this latter category. Therefore, they
cannot be used as arguments for any real phenomenon; in particular,
for the phenomenon of the EoS.

Still, it is instructive to understand why these effects appear in the
RW coordinates. The reason is as follows. In cosmology, we usually
work in the RW coordinates, or in `public space': we use local
coordinates of all FOs, and in particular we measure common `cosmic
time'. However, an alternative approach is possible: an analysis in
`private space' of a given FO, employing his `private time'. In the
Milne model (empty universe), private space and time are defined by
the Minkowskian coordinates of the FO's inertial rest-frame. In
non-empty models, they are defined by a natural generalization thereof
(i.e., conformally Minkowskian coordinates). Public space is a hybrid
of many local `private spaces', or inertial rest-frames of different
FOs, all in relative motion. As a result, public space is not a global
inertial frame, {\em even in the Milne model,} while the velocity of
light in vacuum is $c$ only in inertial frames. More specifically, the
time and length measures of different FOs are subject to relativistic
time-dilation and length-contraction. It is therefore not surprising
that in public space, the superluminality of distant galaxies and the
travel-time effect are present even in the Milne model. Like other FL
models, this model possesses also `acausal' distance to the particle
horizon (greater -- in fact, infinitely -- than $c\, \tau_0$, where
$\tau_0$ is the age of the universe). The explanation is again
time-dilation (Chodorowski 2007), and all FL models with no initial
period of deceleration are expected to possess no particle horizon
(or, to have it infinite).

We work in public space, or RW coordinates, for convenience. The
symmetries, which we endow our model of the universe with, are
apparent in the resulting field (FL) equations. In particular, the
density field of the cosmic substratum is homogeneous and isotropic
everywhere. In the conformally Minkowskian coordinates this is no
longer true: the density field around a given FO is isotropic but not
homogeneous (Landau \& Lifshitz 1979). The RW coordinates are thus
more convenient for calculations. However, the conformally Minkowskian
coordinates are useful to interpret some of the results, obtained in
RW coordinates, which defy our expectations based on SR. As stated
above, among them are the superluminality of distant galaxies, acausal
distance to the horizon and the travel-time effect. A `common
denominator' of all these effects is relativistic time dilation. This
physical explanation contrasts with an alternative, standard
explanation: a consequence of the EoS. In the latter case, the EoS
serves effectively as a rug under which we sweep up everything we
don't fully understand.

\section{Summary and concluding remarks}
\label{sec:summ}
This paper has been devoted to a critical discussion of the concept of
the Expansion of Space (EoS) in cosmology. We have argued that
expanding space is as real as ether, in a sense that they are both
unobservable. More specifically, propagation of light is a
relativistic phenomenon: for light, the analogy of a swimmer in a
river does not work; the velocity of light is $c$ in every inertial
frame. This explains the null results of all the ether-drift
experiments, and enables one to predict the null results of any
expanding -- or drifting -- space experiments.

We have shown that both the superluminality of distant galaxies and
the travel-time effect for photons are {\em merely coordinate
effects}: they vanish in a suitably chosen coordinate system.
Therefore, they are not real phenomena, which different observers will
agree on. In the Milne model, the travel-time effect -- present in the
RW coordinates -- is explicable entirely by the relativistic
phenomenon of time dilation. Since in the real universe distant
galaxies recede with relativistic velocities, time dilation must play
a role also in the case of more realistic FL models.

The concept of the EoS has been invented to stress that the GR
description of the expansion of the universe can conflict with our
intuitions based on SR. However, for non-specialists this concept can
be very misleading: in their minds, it can easily become endowed with
force or some sort of physical or causal power. This point has been
extensively discussed in Section~\ref{sec:intro}. Therefore, the
author of the present paper prefers to advocate an alternative,
semi-popular description, or model, of the universe and its expansion.
Namely, the universe is like the Milne model, but with effects of
mutual gravity. Gravity modifies relative motions of the particles of
the cosmic substratum and makes GR in cosmology indispensable. The
conflict of the GR description of distant events in the universe with
our SR expectations is only apparent: the velocity of light in vacuum
is $c$ only in inertial frames, while in the real universe such frames
are only of limited extent.

Is the concept of the EoS dangerous also for specialists? Not
necessarily. Some specialists use it, but in a somewhat different
sense: for them, the EoS is just the GR solution for the expansion of
the universe when expressed in RW coordinates (Davis, private
communication). Also, all relativists agree that matter and space are
inexorably intertwined in GR. Therefore, indeed the debate on the
meaning and the use of the phrase `Expansion of Space' ``is somewhat a
matter of philosophy and semantics, rather than hard science'' (Davis,
private communication). However, we believe that philosophy and
semantics do matter in cosmology. Therefore, we suggest to avoid using
the phrase `Expansion of Space', as potentially leading to confusion
and wrong intuitions.

\section*{Acknowledgments}
This paper originated as a result of numerous stimulating discussions
with Krzysztof Bolejko. I acknowledge also inspiring discussions with
Frank Rubin. I am grateful to Pawe{\l} Nurowski and Andrzej Trautman
for calling my attention to the paper of Infeld and Schild. The
referee, Tamara Davis, is acknowledged for a very constructive
critisism and many apt comments. This research has been supported in
part by the Polish State Committee for Scientific Research grant No.~1
P03D 012 26, allocated for the period 2004--2007.

\end{document}